\documentclass[12pt]{article}

\usepackage{geometry}
\usepackage{subfig}
\usepackage{rotating}
\usepackage{array}
\usepackage[doublespacing]{setspace}
\usepackage[page]{appendix}
\usepackage{xspace}
\usepackage{parskip}

\usepackage{amsmath}
\usepackage{verbatim}
\usepackage{caption}
\usepackage{graphicx}
\usepackage[sc]{mathpazo}
\usepackage{supertabular}

\newcommand{\Post}{\pi(\theta|y)}
\newcommand{\LikePrior}{f(y|\theta)\pi(\theta)}

\newcommand{\Gtheta}{g(\theta)}
\newcommand{\Ftheta}{\Phi(\theta)}
\newcommand{\Phitheta}{\Phi(\theta|y)}
\newcommand{\Ly}{\mathcal{L}(y)}
\newcommand{\Dy}{\mathcal{D}(\theta,y)}

\DeclareGraphicsExtensions{.png,.pdf}

\begin{document}

\title{Generalized Direct Sampling for Hierarchical Bayesian Models}
\author{Michael Braun\\
MIT Sloan School of Management\\
Massachusetts Institute of Technology\\
Cambridge, MA 02142\\
braunm@mit.edu
\and
Paul Damien\\
McCombs School of Business\\
University of Texas at Austin\\
Austin, TX 78712\\
paul.damien@mccombs.utexas.edu
}
\date{August 7, 2012}

\begin{singlespace}
\maketitle
\begin{abstract}

We develop a new method to sample from posterior
distributions in hierarchical models without using Markov chain Monte
Carlo. This method, which is a variant of importance sampling ideas,
is generally applicable to high-dimensional models involving large
data sets. Samples are independent, so they can be collected in
parallel, and we do not need to be concerned with issues like chain
convergence and autocorrelation.  Additionally, the method can be used to compute marginal likelihoods.

\end{abstract}
\end{singlespace}
\thispagestyle{empty}
\newpage
\setcounter{page}{1}

\section{Introduction}

Recently, Walker et. al (2010) introduced and demonstrated the merits
of a non-MCMC approach called Direct Sampling (DS) for conducting
Bayesian inference.  They argued that with their method there is no need to concern oneself with issues like chain convergence and autocorrelation. They also point out
that their method generates independent samples
from a target posterior distribution \emph{in parallel}, unlike MCMC
for which, in the absence of parallel independent
chains, samples are collected sequentially. Walker et al. also prove that the sample acceptance probabilities using DS are better than those from standard 
rejection algorithms.  Put simply, for many common Bayesian models, they demonstrate 
improvement over MCMC in terms of its efficiency, resource demands and
ease of implementation.  

However, DS
suffers from some important shortcomings that limit its broad
applicability. One is the failure to
separate the specification of the prior from the specifics of the
estimation algorithm.  Another is an inability to generate accepted draws for
even moderately sized problems; the largest number of parameters that
Walker et al. consider is 10.  Our interest is
in conducting full Bayesian inference on hierarchical models in high dimensions,
with or without conjugacy, sans MCMC.  

The method proposed in this paper, strictly speaking, is not a generalization of the DS algorithm, but since it shares some important features with DS, we call it Generalized Direct Sampling (GDS). DS and GDS differ in the following respects.
\begin{enumerate}
\item  While DS focuses on the shape of the data likelihood alone,
  GDS is concerned with the characteristics of the entire posterior density.
\item GDS bypasses the need for Bernstein polynomial approximations,
  which are integral to the DS algorithm.
\item While DS takes proposal draws from the prior (which may conflict
  with the data), GDS samples
  proposals from a separate density that is ideally a good
  approximation to the target posterior density itself.
\end{enumerate}

In addition to the above improvements over DS, GDS maintains many
improvements over MCMC estimation:
\begin{enumerate}
\item All samples are collected independently, so there is no need to be
  concerned with autocorrelation, convergence of estimations chains,
  and so forth.
\item There is no particular advantage to choosing model components
  that maintain conditional conjugacy, as is common with Gibbs sampling.
\item GDS generates samples from the target posterior entirely in
  parallel, which takes advantage of the most recent advances in
  grid computing and placing multiple CPU cores in a single
  computer.  
\item GDS permits fast and accurate estimation of marginal likelihoods
  of the data.
\end{enumerate}

GDS is introduced in Section \ref{sec:GDS}.  Section \ref{sec:empirical} includes examples of GDS in action, starting with a small, but important,
two-parameter example for which MCMC is known to fail, and concluding
with a complex nonconjugate application with over 29,000 parameters.
In Section \ref{sec:MargLL}, GDS is used to estimate
marginal likelihoods. Finally, in Section \ref{sec:practical}, we
discuss practical issues that one should consider when
implementing GDS, including limitations of the approach.

\section{Generalized Direct Sampling:  The Method}\label{sec:GDS}
Like DS, GDS is a variant on well-known importance sampling methods. The goal is to
sample $\theta\in\Omega$ from a posterior density
\begin{align}
  \label{eq:Post}
  \Post=\frac{\LikePrior}{\Ly}=\frac{\Dy}{\Ly}
\end{align}
where $\Dy$ is the joint
density of the data and the parameters (the unnormalized posterior density).  Let $\theta^*$ be the mode of
$\Dy$, and define $c_1=\mathcal{D}(\theta^*,y)$. Choose some proposal
distribution $\Gtheta$ that also has its mode at $\theta^*$, and
define $c_2=g(\theta^*)$.  Also, define the function
\begin{align}
  \label{eq:PhiDef}
\Phitheta=\displaystyle\frac{\LikePrior \cdot c_2}{\Gtheta \cdot c_1}  
\end{align}
 Obviously
\begin{align}
  \label{eq:PostDefPhi}
  \Post=\Phitheta\cdot\Gtheta\cdot\frac{c_1}{c_2\cdot\Ly}
\end{align}
An important restriction on the choice of $g(\theta)$ is that the inequality $0<\Phi(\theta|y)\leq 1$ must hold, at least for any
$\theta$ with a non-negligible posterior density.  
Discussion of the choice of $\Gtheta$ is given in detail a little later.

Next, let $u|\theta,y$ be an auxiliary variable that is distributed uniformly on
$\left( 0,\displaystyle\frac{\Phitheta}{\pi(\theta|y)}\right)$, so that
$p(u|\theta, y)=\displaystyle\frac{\pi(\theta|y)}{\Phitheta}=\frac{c_1}{c_2\Ly}g(\theta)$.  We then construct a joint density of
$\theta|y$ and $u|\theta, y$, where
\begin{align}\label{eq:joint}
 p(\theta,u|y)&=\frac{\pi(\theta|y)}{\Phitheta}\mathbb{1}\left[ u<\Phitheta\right] 
\end{align}
From Equation \ref{eq:joint}, the marginal density of $\theta|y$ is
\begin{align}
  \label{eq:margThetaY}
  p(\theta|y)&=\frac{\pi(\theta|y)}{\Phitheta}\int_0^{\Phitheta}~du =\pi(\theta|y)
\end{align}
Therefore, simulating
from $p(\theta|y)$ is equivalent to simulating from the target posterior $\pi(\theta|y)$.

Using Equations \ref{eq:PostDefPhi} and \ref{eq:joint}, the marginal density of $u|y$ is
\begin{align}
  \label{eq:margU}
p(u|y)&=\int \frac{\pi(\theta|y)}{\Phi(\theta|y)}\mathbb{1}\left[
  u<\Phi(\theta|y)\right]  d\theta\\
&=\frac{c_1}{c_2\Ly}\int \mathbb{1}\left[
  u<\Phi(\theta|y)\right] g(\theta)~d\theta\\
&=\frac{c_1}{c_2\Ly}q(u)\label{eq:marg_uy}
\end{align}
where $q(u) =\int \mathbb{1}\left[
  u<\Phi(\theta|y)\right] g(\theta)~d\theta$ is defined as the probability that
$u<\Phitheta$ for any $\theta$ drawn from $\Gtheta$.

The GDS sampler comes from recognizing that $p(\theta,u|y)$ can
written differently from, but equivalently to, Equation \ref{eq:joint}.
\begin{align}
  \label{eq:joint2}
  p(\theta,u|y)&=p(\theta|u,y)~p(u|y)
\end{align}
The strategy behind GDS is to sample from an approximation to $p(u|y)$, and then sample from
$p(\theta|u,y)$.  Using the definitions in Equations \ref{eq:PhiDef},
\ref{eq:PostDefPhi}, and \ref{eq:joint}, we get
\begin{align}
  \label{eq:GDS_AR}
  p(\theta|u,y)&=\frac{p(\theta,u|y)}{p(u|y)}\\
&=\frac{c_1}{c_2\Ly}\frac{\mathbb{1}[u<\Phitheta]~g(\theta)}{p(u|y)}
\end{align}
Consequently, to sample \emph{directly} from $\pi(\theta|y)$, one
needs only to sample from $p(u|y)$ and then sample repeatedly from
$g(\theta)$ until $\Phi(\theta|y)>u$.

How does one simulate from $p(u|y)$, which is proportional to $q(u)$?
Walker et al. sample from a similar kind of density by first taking $M$ proposal draws from the prior to construct
an empirical approximation to $q(u)$, and then constructing a
continuous approximation using Bernstein polynomials.  However, in
high-dimensional models, this approximation tends to be a poor one at
the endpoints, even with an extremely large number of Bernstein polynomial
components. It is for this reason that the largest number of
parameters that Walker et. al. tackle is 10.  

The GDS strategy to sample $u|y$ is to sample a transformed variable
$v=T(u)$, where $T(u) = -\log u$.  Applying a change of variables,
$q(v)=q(u)\exp(-v)$.  With $q(v)$ denoting the ``true'' CDF of $v$,
let $q_M(v)$ be the empirical CDF after taking $M$ proposal draws from
$\Gtheta$, and ordering the proposals
$0<v_1<v_2<\mathellipsis<v_M<\infty$. To be clear,
$q_M(v)$ is the proportion of proposal draws that are \emph{strictly}
less than $v$.  Because $q_M(v)$ is discrete, we can sample from a
density proportional to $q(v)\exp(-v)$ by partitioning the domain into
$M+1$ segments, partitioned at each $v_i$.  The probability of
sampling a new $v$ that falls between $v_i$ and $v_{i+1}$ is now
$\varpi_i=q_M(v)\left[\exp(-v_i)-\exp(-v_{i+1})\right]$.  Therefore, we
first sample a $v_i$ from a multinomial density with weights proportional to $\varpi_i$, 
and then let $v=v_i + \epsilon$, where $\epsilon$ is a draw from a standard
exponential density, truncated on the right at $v_{i+1}-v_i$. One can
sample from this truncated exponential density by first
sampling a standard uniform random variate $\eta$, and setting $v=-\log\left[1-\eta\left(1-\exp(v_i-v_{i+1}\right)\right]$.

To sample $N$ draws from the target posterior, we sample $N$ ``threshold''
draws of $v$ using this method.  Then, for each $v$, we repeatedly
sample from $\Gtheta$ until  $T(\Phitheta)< v$.  Note that the
inequality sign is flipped from the $\Phitheta>u$ expression because of the negative sign in the transformation.

In summary, the steps of the GDS algorithm are as follows:

\begin{enumerate}
\item Find the mode of $\Dy$, $\theta^*$ and compute the
  unnormalized log posterior
  density $c_1=\mathcal{D}(\theta^*,y)$ at that mode.
\item Choose a distribution $g(\theta)$ so that its mode is also at
  $\theta^*$, and let $c_2=g(\theta^*)$.
\item Sample $\theta_{1},\ldots,\theta_{M}$ independently from
  $g(\theta)$.  Compute $\Phi(\theta_m)$ for these
  proposal draws.  If $\Phi(\theta_m)>1$ for any of these draws, repeat Step
  2 and choose another proposal distribution for which $\Phi(\theta_m)<
  1$ does hold.
\item Compute $v_i=T(\Phi(\theta|y))$ for the $M$ proposal draws, and place
  them in increasing order.
\item Evaluate, for each proposal draw, 
\begin{align}
 q_{M}(v)=\sum_{i=1}^M\mathbb{1}\left[v_i<v\right]
\end{align}
which is the empirical CDF of $v_i$ for the $M$ proposal draws. Then
compute $\varpi_i=q_M(v)\left[\exp(-v_i)-\exp(-v_{i+1})\right]$ for
all $i$.
\item  Sample $N$ draws of $v=v_i+\epsilon$, where a particular $v_i$ is chosen
  according to the multinomial distribution with probabilities
  proportional to $\varpi_i$, and
  $\epsilon$ is a standard exponential random variate, truncated to $v_{i+1}-v_i$.
\item For each of the $N$ required samples from the target posterior, sample $\theta$ from $g(\theta)$ until
  $T(\Phitheta)< v$.  Consider
  each first accepted draw to be a single draw from the target
  posterior $\pi(\theta|y)$.
\end{enumerate}

Choosing $\Gtheta$ is an important part
of this algorithm.  Naturally, the closer $\Gtheta$ is to the target
posterior, the more efficient the algorithm will be in terms of
acceptance rates. In principle, it is up to the researcher to choose
$\Gtheta$, which is similar in spirit to selecting a dominating density
while implementing standard rejection algorithms, or even
Metropolis-Hastings algorithms.  For GDS, in practice, a multivariate
normal proposal distribution with mean at $\theta^*$ and
covariance matrix of the inverse Hessian at $\theta^*$, multiplied by a
scaling constant $s$, works well. There is nothing special about this 
choice, except to note that it is easy to implement with a little
trial and error in the selection of $s$. (This is similar, in spirit, to the 
concept of tuning an M-H algorithm via trial and error.) If the log posterior happens
to be multimodal, and the location of the local modes are known, then
one could let $\Gtheta$ be a mixture of multivariate normals instead. Importantly, we
address the sensitivity of the GDS algorithm to $M$ as part of the
analysis in Section \ref{sec:MargLL}.

Clearly, an advantage of GDS is that the samples one collects from
the target posterior density are \emph{independent}, and that lets us
collect them in parallel. Some researchers have investigated alternative
approaches for MCMC-based Bayesian inference that also take advantage
of parallel computation; see, for example, Suchard et al. (2010).  One
notable example is a parallel implementation of a multivariate slice
sampler (MSS), as in Tibbits et. al. (2010).   The benefits of parallelizing the MSS
come from parallel evaluation of the target density at each of the vertices of the
multivariate slice, and from more efficient use of resources to execute
linear algebra operations (e.g, Cholesky decompositions). But the MSS
itself remains a Markovian algorithm, and thus will still generate dependent
draws.  Using parallel technology to generate a single draw from a distribution is not the
same as generating all of the required draws themselves in parallel.
On the other hand, the sampling steps of GDS can be run in their
\emph{entirety} in parallel.  

\section{Illustrative Analysis}\label{sec:empirical}

We now provide some examples of GDS in action, especially on problems for which
MCMC fails, or for which the dimensionality, model
structure, and sample size make MCMC methods somewhat
unattractive. 

\subsection{A Hierarchical Non-Gaussian Linear Model}\label{sec:cauchy}

\begin{figure}[tbp]
    \centering
    \includegraphics[scale=.5]{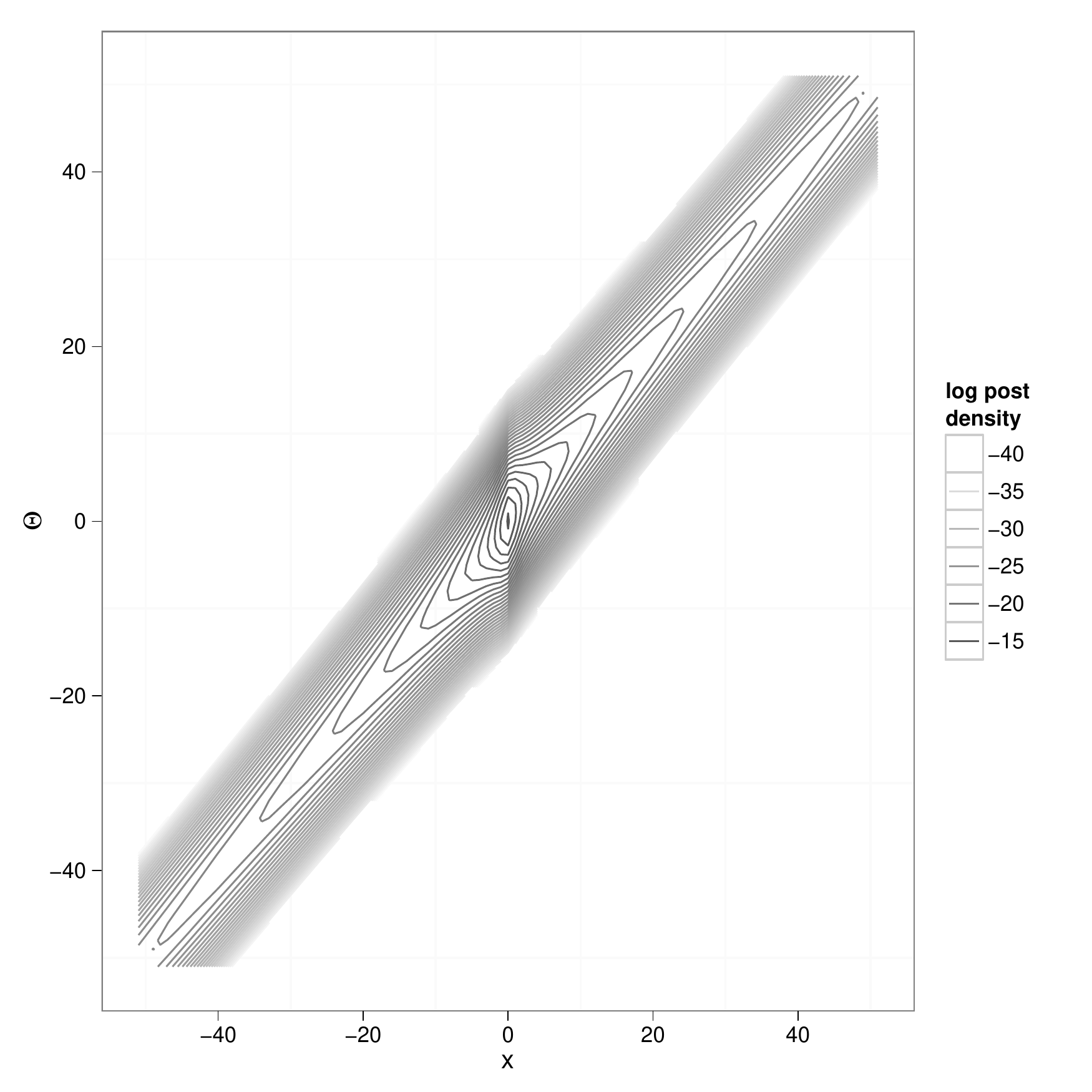}
    \caption{\small Contours of the "true" posterior distribution of the non-Gaussian linear model example.}
    \label{fig:cauchyContours}
  \end{figure}

Consider this motivating example of a linear hierarchical model discussed by
Papaspiliopoulous and Roberts (2008). 
\begin{align}\label{eq:2}
  Y&=X + \epsilon_1\\
X&=\Theta + \epsilon_2
\end{align}
For an observed value $Y$ , $X$ is the
latent mean for the prior on $Y$, $\Theta$ is the prior mean of $X$, and $\epsilon_1$ and $\epsilon_2$ are random error terms, each with mean
0.  Papaspiliopoulous and Roberts note that to improve the robustness of inference on $X$ to
outliers of $Y$, it is common to model $\epsilon_1$ as having  heavier tails than
$\epsilon_2$ .   Let $\epsilon_1 \sim \text{Cauchy}(0,1)$, $\epsilon_2 \sim
N(0,5)$, and  $\Theta \sim N(0, 50000)$, and suppose there is only one observation available, $Y=0$.
The posterior joint distribution of $X$ and $\Theta$ is given in Figure
\ref{fig:cauchyContours}; the contours represent the logs of the
computed posterior densities.  Note that around the
mode, $X$ and $\Theta$ appear uncorrelated, but in the tails they are
highly dependent.  Papaspiliopoulous and Roberts present this example
as a deceptively simple case in which Gibbs sampling performs extremely
poorly. Indeed, they note that almost all diagnostic tests will erroneously
conclude that the chain has converged. The reason for this failure is that the MCMC
chains are attracted to, and get ``stuck'' in, the modal region where the variables are
uncorrelated.  Once the chain enters the tails, where the variables
are more correlated, the chains moves slowly, or not at all.

GDS is a more effective alternative for sampling from the posterior distribution.  The posterior mode and
Hessian of the log posterior at the mode, are $\theta^*=(0, 0)$
and $ H = \left(
\begin{array}{cc}
-2.2 & 0.2 \\
 0.2 & -0.2
\end{array} \right)$.  The GDS proposal distribution $\Gtheta$ is taken to be a bivariate normal with mean
$\theta^*$ and covariance $-sH^{-1}$, with $s=200$.  This scaling
factor was the smallest value of $s$ for which $\Ftheta\leq 1$ for all $M=20,000$ of the
proposal draws.  Two hundred independent samples were collected using the
GDS algorithm.

\begin{figure}[tbp]
  \centering
  \includegraphics[scale=.5]{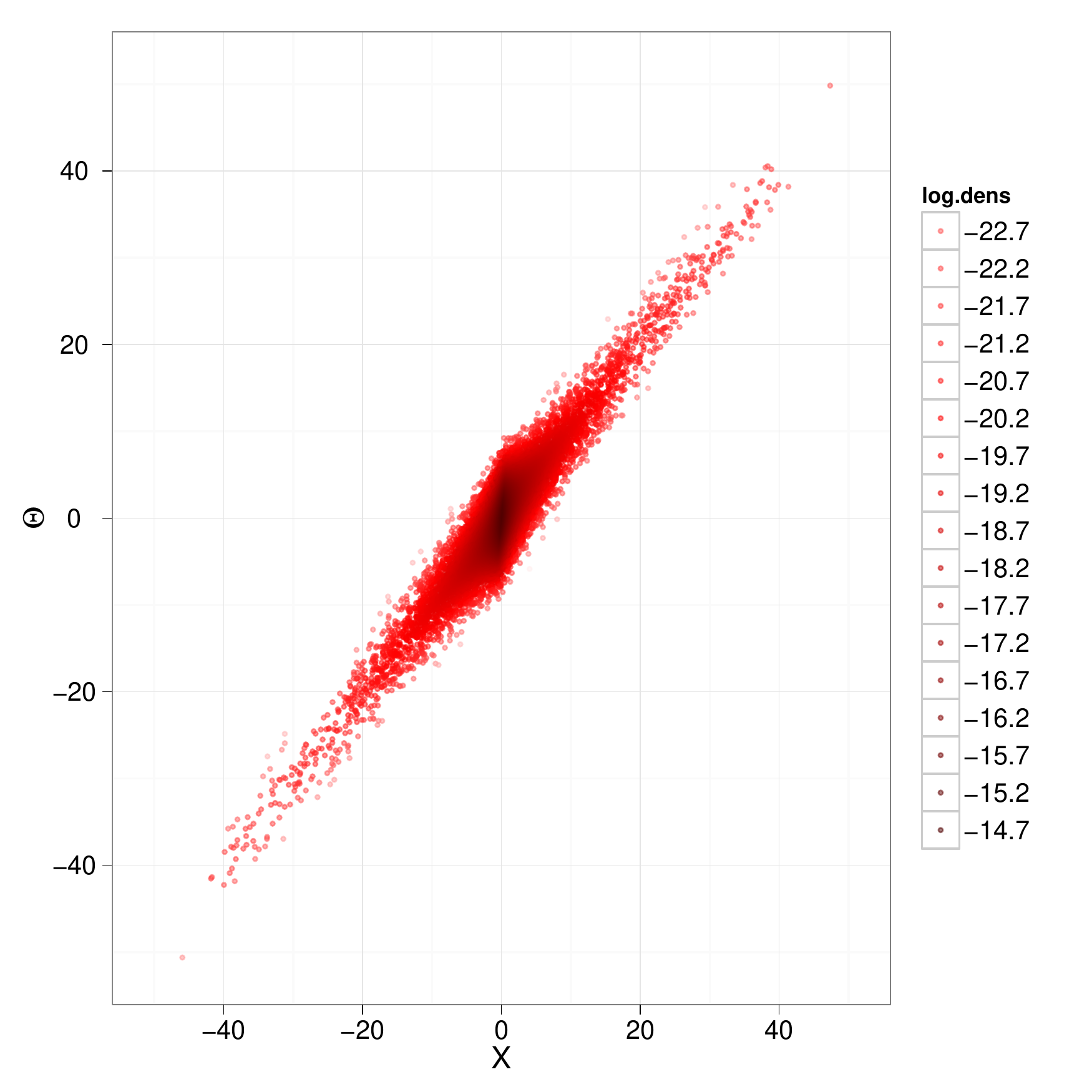}
  \caption{\small Posterior draws from non-Gaussian linear regression
    example, using GDS.  Darker colors represent regions of higher posterior density}
  \label{fig:cauchyDraws}
\end{figure}

 Figure \ref{fig:cauchyDraws} plots each of the GDS draws,
where darker areas represent higher values of log posterior density.  GDS not only picks up the correct shape of the regions of
high posterior mass near the origin, but also the dependence in the
long tails.  The acceptance rate to collect these draws was about
0.013.

\begin{figure}[tbp]
  \centering
  \includegraphics[scale=.5]{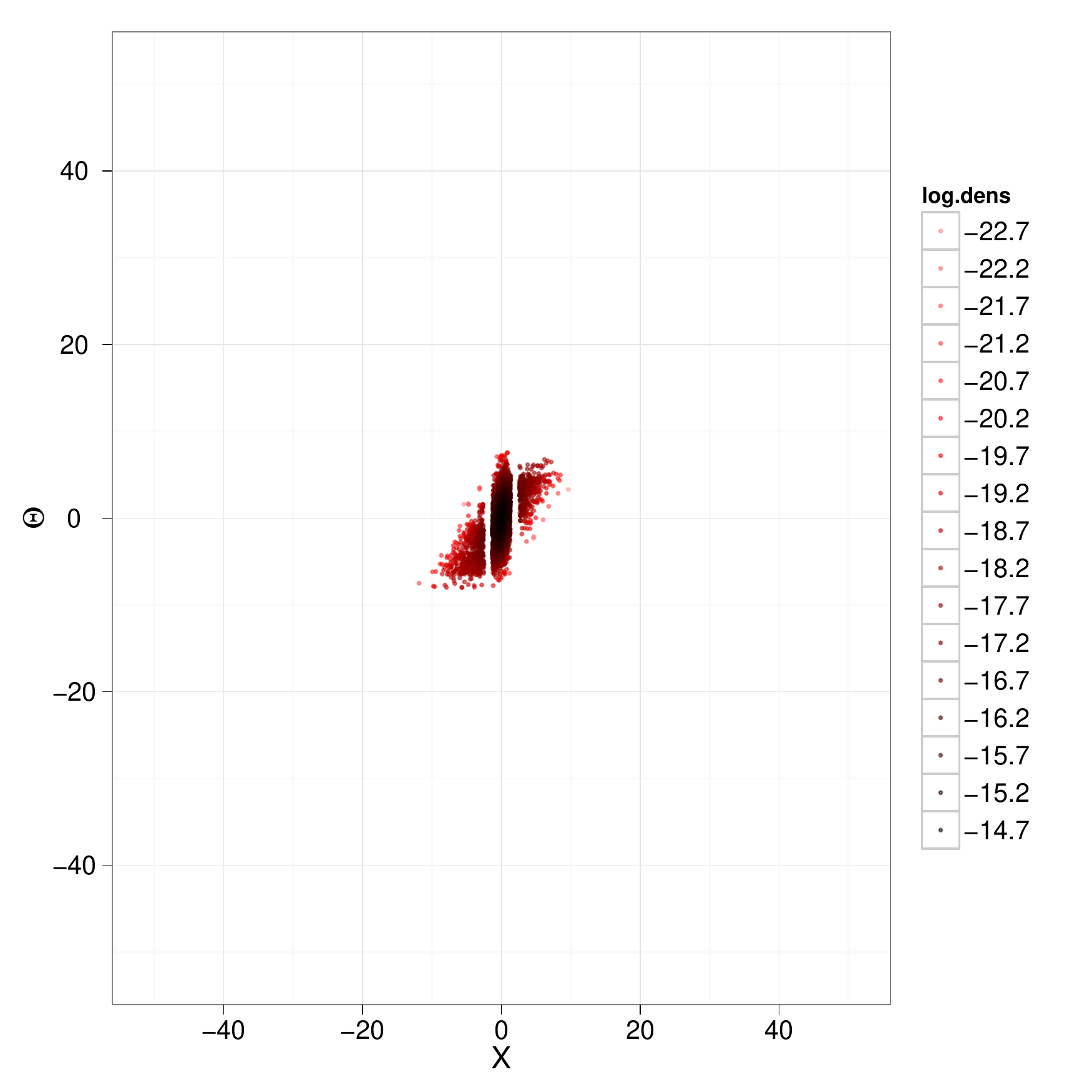}
  \caption{\small Posterior draws from non-Gaussian linear regression
    example, using MCMC.  Darker colors represent regions of higher posterior density}
  \label{fig:cauchyMH}
\end{figure}

In contrast, consider Figure \ref{fig:cauchyMH}. which plots samples collected using
MCMC.  Specifically, we used the RH-MALA method in Girolami and
Calderhead (2012), with constant curvature, estimating the Hessian at
each iteration. These are samples from 25,000 iterations, collected after
starting at the posterior mode and running through 25,000 burn-in
iterations, thinned every 10 draws.  Just as Papaspiliopolous and Roberts predicted, the
chain tends to get stuck near the mode. It is only after some
serendipitously large proposal jumps that the chain ever finds itself
in the tails (hence the gaps in the plot), but the chain does not move
very far along those tails at all.

\subsection{Hierarchical Gaussian model}\label{sec:HierMVN}

Next, we consider a hierarchical model with a large number of
parameters.  The dependent variable $y_{it}$ is measured $T$ times for
heterogenous units $i=1\mathellipsis n$.  For each unit, there are $k$
covariates, including an intercept. The intercept and coefficients
$\beta_i$ are heterogeneous, with a Gaussian prior with mean
$\bar{\beta}$ and covariance $\Omega$, which in turn have weakly
information standard hyperpriors.  This model structure is given by:

\begin{align}
  y_{it}&\sim N(x_i'\beta_i, 1),~i=1\mathellipsis n,~t=1\mathellipsis T\\
\beta_i|\Omega&\sim MVN(\bar{\beta}, \Omega)\\
\bar{\beta}&\sim MVN(0,V_{\beta})\\
\Omega&\sim IW(\nu,A)
\end{align}

To construct simulated datasets, we set ``true'' values of
$\bar{\beta}=(5,0,-2,0)$ and $\Omega=0.25I$.  For each unit, we
simulated $T=25$ observations, where the non-intercept
covariates are all i.i.d draws from a standard normal distribution.
Three different values for $n$ were entertained:  100, 500 and 1000.   In all cases,
there are 14 population-level parameters, so the total number of
parameters are 414, 2014 and 4014, respectively.  The parameters of
the hyperpriors are $\nu=10$, $V_{\beta}=0.2I$, and
$A=0.1I$. 

Table \ref{tab:HierMVN_gds} summarizes the performance
of the GDS algorithm, averaged over 10 replications of the
experiment.  For
each case, we collected 100 samples from the posterior, with
$M=10,000$ proposal draws.  The proposal distributions are all multivariate normal,
with mean at the posterior mode and the covariance set as the inverse
of the Hessian at the mode, multiplied
by a scale factor that changes with $n$.  For each value of $n$,  we
set the scale factor to roughly be the
smallest value for which $\Phitheta\leq 1$ for all $M=10,000$ proposal
draws.  The ``Mean Proposals'' column is the average number of
proposals it took to collect 100 posterior draws during the
accept-reject phase of the algorithm (this is the inverse of the
acceptance rate).  We also recorded the time it took to run each stage
of the algorithm.  The ``Post Mode'' column is the number of minutes it
took to find the posterior mode, starting at the origin (after
transforming all parameters to have a domain on the real line).  The
``Proposals'' column is the amount of time it took to collect the  $M=10,000$
proposal draws.  ``Acc-Rej'' is the time it took to execute the
accept-reject phase of the algorithm to collect 100 independent
samples from the posterior,
and ``Total'' is the total time required to run the algorithm.  The study
was conducted on an Apple Mac Pro with 12 CPU cores running at 2.93GHz
and 32GB of RAM.  Ten of the 12 cores were allocated to this algorithm.

\begin{table}[h]
  \centering
  \begin{tabular}[h]{rrr|rrrrr}
&Total&scale&Mean&\multicolumn{4}{c}{Minutes for GDS stages}\\
$n$&params&factor&Proposals&Post Mode&Proposals&Acc-Rej&Total\\
\hline
100 & 414 & 1.32 & 15489 & 0.01 & 0.11 & 1.9 & 2.1 \\ 
  500 & 2014 & 1.20 & 23387 & 0.08 & 0.18 & 11.3 & 11.6 \\ 
  1000 & 4014 & 1.14 & 26480 & 0.24 & 0.28 & 23.0 & 23.6 \\ 
  \end{tabular}
  \caption{Efficiency of GDS for hierarchical Gaussian example}\label{tab:HierMVN_gds}
\end{table}

Although Mean Proposals may appear to be high, assessing the efficiency of an algorithm by examining the
acceptance rate alone can be misleading.  For a single estimation
chain, MCMC cannot be run in parallel, while for GDS, all draws can be
generated in parallel.  For example, each of the 10 cores on the CPU
was responsible for collecting 10 posterior samples.  One possible counterargument would be that
MCMC can be run with multiple chains in parallel, as suggested in
Gelman and Rubin (1992).  However, if one does use parallel chains,
all of the chains need to be burned in independently. There is no
guarantee that any one of the chains would have converged after some arbitrary
number of iterations, and there will still be residual interdependence
in the final set of collected draws.  It would be more appropriate to compare the
total number of proposals from the GDS rejection sampling phase to the
number of post-burn-in iterations that MCMC runs require reaching an
\emph{effective sample size} of 100.  It is also worth reiterating that GDS is quite simple to
implement, as the algorithm requires no ongoing tuning or adaptation
once the accept-reject phase begins. 

As a point of comparison, we estimated the posterior for the $n=500$ case using the
adaptive Metropolis-adjusted Langevin algorithm with truncated drift
(Atchade 2006). We chose this algorithm because it is commonly used,
not too hard to implement (relative to many other MCMC variants lie
Gibbs sampling), generically applicable to a general class of models
that are similar to those for which GDS would work well, and
exploits the gradient information that GDS uses to find the posterior
mode. We started the chain
at the posterior mode, and let it run for 5.5 million iterations,
during the course of several weeks.  Using the Geweke convergence
diagnostic criterion on the log posterior density (Geweke 1991), we decided that only the final 180,000 draws
could be considered ``post burn-in.'' However, those draws represented
an effective sample size of only 1,460, suggesting a required thinning
ratio of about 1 in 12,000. We did try some more advanced algorithms, such as
  Girolami and Calderhead(2012), but had difficulty when the chain
  entered regions for which the Hessian was not negative definite.
  The Atchade algorithm generates updates of the proposal covariance
  that are guaranteed to be positive definite. Regardless, this case illustrates how unreliable (and
unpredictable) MCMC can be for collecting posterior samples quickly and easily.

\subsection{An example of a complex, high-dimensional model}\label{sec:ads}

In this
section, we consider another model for which GDS should be an
attractive estimation method:  the effectiveness of online advertising
campaigns.  Using GDS, we estimate the posterior density from the
model described in Braun and Moe (2012).  This model is quite
complicated; for 5,803
anonymous users, they observe which website advertisements (if any) 
were served to each user during the course of that user's web browsing
activity.  They also observe when these users visited the
advertiser's own website (if ever), and if these website visits
resulted in a conversion to a sale.
The managerial objective is to develop a
method for firms to identify which versions of ads are most likely to generate
site visits and sales, taking into account the fact that the return on
investment of the ad may not occur until several weeks in the future.
The model allows each version of an ad to have a contemporaneous effect in that week, but
for each repeat view of the same ad to an incrementally smaller effect.
The effect of the ad campaign for an individual builds up with each subsequent
ad impression, but this accumulated ``ad stock'' decays from week to
week. 

All together, there are 29,073 parameters in the model, consisting of
five heterogeneous parameters per user, plus 58 population-level
parameters. The data are modeled as being generated from zero-inflated
Poisson (for ads and visits) and binomial (for successes)
distributions, with the rate parameters being correlated, and
dependent on a latent ad stock variable.  The ad stock, in turn, is
depends on which version of the ad is served, along with nonlinear
representations of build-up, wear-out and restoration effects.  Given
the complex hierarchical structure, extensive nonlinearities, and high
degree of correlation in the posterior distribution, attempts at MCMC
estimation proved to be well-nigh impossible to implement.

GDS, however, worked well.  Because the user-level data are
conditionally independent, we could write the log posterior density as
the sum of user-level data likelihoods, plus the priors and
hyperpriors.  Algorithmic differentiation software (Bell 2012)
automatically computed the gradients and Hessians for the model. This
allows us to find the posterior mode, and estimate
the Hessian at the mode, relatively quickly.  As above, the proposal
density was a multivariate normal, centered at the posterior mode,
with a covariance matrix of 1.02 times the inverse of the Hessian at
the model.  Since the model assumes conditional independence across
users, the Hessian of the log posterior has a sparse structure.  We
exploit this sparsity to generate the proposal draws efficiently, and
to dramatically reduce the memory footprint of the algorithm.

To collect 100 independent draws, it took 23.75 minutes.  The mean attempts per draw is 2350.
  Although this may appear like a low acceptance rate, consider that we were
able to collect the draws in parallel, without any tuning or
adaptation of the algorithm beyond the initial choice of the proposal density.  Also, because the sparse Hessian
has a ``block-diagonal-arrow'' structure, it grows only linearly
with the number of users.

\section{Estimating Marginal Likelihoods}\label{sec:MargLL}
Now, we turn to another advantage of GDS:  the
ability to generate accurate estimates of the marginal
likelihood of the data with little additional computation.  A number of researchers have proposed methods to
approximate the marginal likelihood, $\Ly$, from MCMC output.  Popular
examples include Gelfand and Dey (1994), Newton and Raftery (1994),
Chib (1995) and Raftery, et. al. (2007).  But none of these methods
have achieved universal acceptance as being unbiased, stable or easy
to compute.  In fact, Lenk (2009) demonstrated that methods which 
depend solely on samples from the posterior density could suffer from a
``pseudo-bias,'' and he proposes a method to correct for it.  Through several examples, he demonstrates that
his method dominates other popular methods, although with substantial
computational effort.  Thus, the estimation of the marginal likelihood of a
dataset, and its use in model selection, remains a 
difficult problem in MCMC-based Bayesian statistics for which there is no satisfactory solution.

To estimate the marginal likelihood from GDS output, we need
the acceptance rate from the accept-reject stage of the algorithm.  Recall that $q(u)$ is the probability that, given a threshold value
$u$, a proposal from $\Gtheta$ is accepted.  Therefore, one can express
the expected marginal acceptance probability for any posterior draw as
\begin{align}
  \label{eq:5}
  \gamma&=\int_0^1q(u)p(u|y)~du.
\end{align}

Substituting Equation \ref{eq:margU},
\begin{align}
  \label{eq:6}
  \gamma&=\frac{c_1}{c_2\Ly}\int_0^1q^2(u)~du
\end{align}
Applying the change of variables so $v=-\log u$, and then rearranging terms,
\begin{align}
  \label{eq:Ly}
  \Ly&=-\frac{c_1}{c_2\gamma}\int_0^{\infty}q^2(v)\exp(-v)~dv.
\end{align}

The values for $c_1$ and $c_2$ are immediately available from the GDS
algorithm.  One can estimate $\gamma$ by treating the number of
proposals required to accept a posterior draw as a shifted geometric
random variable (an acceptance on the first proposal is a count of
1).  Thus, an estimator of $\gamma$ is the inverse of the mean number
of proposals per draw.

What remains is estimating the integral in Equation \ref{eq:Ly}.  This is done 
by using the proposal draws from Step 5  in
the GDS algorithm in Section \ref{sec:GDS}.  The
empirical CDF of these draws is discrete, so we can partition the support of $q(v)$
at $v_1\mathellipsis v_M$.  Also, since $q_M(v)$ is the
proportion of proposal draws less than $v$, we have
$q_M(v)=\frac{i}{M}$. Therefore, 
\begin{align}
  \label{eq:9}
  \int_0^{\infty}q^2(v)\exp(-v)dv&\approx\sum_{i=1}^M\int_{v_i}^{v_{i+1}}\left(\frac{i}{M}\right)^2\exp(-v_i)dv\\
&=\frac{1}{M^2}\sum_{i=1}^M i^2\left[\exp(-v_i)-\exp(-v_{i+1})\right]\\
&=\frac{1}{M^2}\sum_{i=1}^M\left( 2i-1\right) \exp(-v_i)
\end{align}

(By convention, define $v_{M+1}=\infty$). Putting all of this together, we can estimate the marginal likelihood as
\begin{align}
  \label{eq:LyEst}
\Ly&\approx\frac{c_1}{M^2c_2\gamma}\sum_{i=1}^M(2i-1)\exp(-v_i)
\end{align}

As a demonstration of the accuracy of this method, consider the following normal linear regression model, also 
used by Lenk (2009) to demonstrate the accuracy of his method.
\begin{align}
  \label{eq:5}
  y_{it}&\sim N(x_i'\beta, \sigma^2),~i=1\mathellipsis n,
  t=1\mathellipsis T\\
\beta|\sigma&\sim N(\beta_0, \sigma^2V_0)\\
\sigma^2&\sim IG(r,\alpha)
\end{align}
For this model, $\Ly$ is a multivariate-T density.  This allows us to compare the estimates of $\Ly$ from GDS
with ``truth.''  To do this, we conducted a simulation study for
simulated datasets of different sizes ($n$=200 or 2000) and numbers of
covariates ($k$=5, 25 or 100).  For each $n,k$ pair, we simulated 25
datasets.  For each dataset, each vector $x_i$ includes an intercept and $k$ iid
samples from a standard normal density.  There are $k+2$
parameters, corresponding to the elements of $\beta$, plus $\sigma$.  The true intercept term is 5, and the remaining true
$\beta$ parameters are linearly spaced from $-5$ to $5$. In all cases,
there are $T=25$ observations per unit.  Hyperpriors are
set as $r=2$, $\alpha=1$, $\beta_0$ as a zero vector and $V_0=0.2\cdot
I_k$.  

For each dataset, we collected 250 draws from the posterior density
using GDS, with different numbers of proposal
draws ($M$=1,000 or 10,000), and scale factors ($s=$0.5, 0.6,
0.7 or 0.8) on the Hessian (so $sH$ is the precision matrix of the MVN
proposal density, and lower scale factors generate more diffuse
proposals).  The $s=0.8$ and $n=200$ cases were excluded because the proposal
density was not sufficiently diffuse that $\Phitheta$ was between 0 and
1 for the $M$ proposal draws.  Table \ref{tab:MargLL1} presents the true log marginal likelihood (MVT), along
with estimates using GDS, the importance sampling method in Lenk
(2009), and the harmonic mean estimator (Newton and Raftery 1994).  Table \ref{tab:MargLL2} shows the mean absolute percentage error
(MAPE) of the GDS and Lenk methods, relative to the true log
likelihood.  The acceptance percentages for each
dataset, and the time it took to generate 250 posterior draws (after finding the
posterior mode) are also summarized.

What we can see from these tables is that the GDS estimates for the log
marginal likelihood are remarkably close to the multivariate T densities, and are 
robust when we use different scale factors. Accuracy
appears to be better for larger datasets than smaller ones, and improving the approximation of $p(u)$ by increasing
the number of proposal draws offers negligible improvement.  Note 
that the performance of the GDS method is comparable to that of
Lenk, but is much better than the harmonic mean estimator.  We did not
compare our method to others (e.g., Gelfand and Dey 1994), because Lenk
already did that when demonstrating the importance of correcting for
pseudo-bias, and how his method dominates many other popular ones.
The GDS method is similar to Lenk's in that it computes the probability
that a proposal draw falls within the support of the posterior
density.  However, note that the inputs to the GDS estimator are
intrinsically generated as the GDS algorithm progresses, while computing
the Lenk estimator requires an additional importance sampling run
after the MCMC draws are collected.  We are not
claiming that our method is better than Lenk's.  Instead,
this illustration shows that one of the important advantages of GDS is the ease and
accuracy with which one can estimate marginal likelihoods.

\begin{table}[hp]\footnotesize
\begin{center}
\begin{tabular}{|llll|rrrrrrrr|}
  \hline
&&&&\multicolumn{2}{c}{MVT}&\multicolumn{2}{c}{GDS}&\multicolumn{2}{c}{Lenk}&\multicolumn{2}{c|}{HME}\\
k & n & M & scale & mean & sd & mean & sd & mean & sd &mean&sd\\
  \hline
 5 & 200 & 1000 & 0.5 & -309 & 6.6 & -309 & 6.6 & -311 & 6.8 & -287 & 7.1 \\ 
   5 & 200 & 1000 & 0.6 & -309 & 6.6 & -309 & 6.7 & -310 & 6.9 & -287 & 6.9 \\ 
   5 & 200 & 1000 & 0.7 & -309 & 6.6 & -309 & 6.7 & -310 & 6.5 & -287 & 6.3 \\ 
    \hline
 5 & 200 & 10000 & 0.5 & -309 & 6.6 & -309 & 6.6 & -311 & 6.7 & -287 & 6.7 \\ 
   5 & 200 & 10000 & 0.6 & -309 & 6.6 & -309 & 6.6 & -310 & 7.5 & -287 & 6.8 \\ 
   5 & 200 & 10000 & 0.7 & -309 & 6.6 & -309 & 6.7 & -310 & 7.0 & -287 & 7.1 \\ 
    \hline
 5 & 2000 & 1000 & 0.5 & -2866 & 46.2 & -2865 & 46.3 & -2868 & 46.2 & -2836 & 46.2 \\ 
   5 & 2000 & 1000 & 0.6 & -2866 & 46.2 & -2866 & 46.2 & -2868 & 45.7 & -2836 & 45.5 \\ 
   5 & 2000 & 1000 & 0.7 & -2866 & 46.2 & -2866 & 46.3 & -2867 & 45.9 & -2836 & 45.9 \\ 
   5 & 2000 & 1000 & 0.8 & -2866 & 46.2 & -2866 & 46.2 & -2867 & 46.3 & -2835 & 46.3 \\ 
   \hline
 5 & 2000 & 10000 & 0.5 & -2866 & 46.2 & -2866 & 46.4 & -2867 & 46.7 & -2836 & 46.9 \\ 
   5 & 2000 & 10000 & 0.6 & -2866 & 46.2 & -2866 & 46.2 & -2867 & 45.8 & -2836 & 46.3 \\ 
   5 & 2000 & 10000 & 0.7 & -2866 & 46.2 & -2866 & 46.4 & -2867 & 46.0 & -2836 & 46.3 \\ 
   5 & 2000 & 10000 & 0.8 & -2866 & 46.2 & -2866 & 46.2 & -2867 & 46.5 & -2835 & 46.3 \\ 
   \hline
25 & 200 & 1000 & 0.5 & -387 & 8.1 & -385 & 8.2 & -391 & 7.6 & -292 & 8.5 \\ 
  25 & 200 & 1000 & 0.6 & -387 & 8.1 & -386 & 8.1 & -390 & 9.5 & -292 & 8.8 \\ 
  25 & 200 & 1000 & 0.7 & -387 & 8.1 & -386 & 8.3 & -390 & 8.0 & -292 & 8.8 \\ 
    \hline
25 & 200 & 10000 & 0.5 & -387 & 8.1 & -385 & 8.5 & -390 & 8.2 & -292 & 8.4 \\ 
  25 & 200 & 10000 & 0.6 & -387 & 8.1 & -385 & 8.2 & -390 & 8.9 & -292 & 8.8 \\ 
  25 & 200 & 10000 & 0.7 & -387 & 8.1 & -386 & 8.2 & -390 & 8.7 & -292 & 9.1 \\ 
    \hline
25 & 2000 & 1000 & 0.5 & -2990 & 28.7 & -2989 & 28.8 & -2994 & 28.3 & -2865 & 28.8 \\ 
  25 & 2000 & 1000 & 0.6 & -2990 & 28.7 & -2989 & 28.7 & -2993 & 28.4 & -2864 & 29.0 \\ 
  25 & 2000 & 1000 & 0.7 & -2990 & 28.7 & -2989 & 28.9 & -2991 & 30.0 & -2864 & 29.5 \\ 
  25 & 2000 & 1000 & 0.8 & -2990 & 28.7 & -2990 & 28.7 & -2992 & 29.6 & -2864 & 29.4 \\ 
   \hline
25 & 2000 & 10000 & 0.5 & -2990 & 28.7 & -2988 & 29.2 & -2992 & 28.5 & -2864 & 28.9 \\ 
  25 & 2000 & 10000 & 0.6 & -2990 & 28.7 & -2989 & 29.1 & -2993 & 29.4 & -2864 & 28.9 \\ 
  25 & 2000 & 10000 & 0.7 & -2990 & 28.7 & -2990 & 29.0 & -2993 & 28.9 & -2864 & 28.9 \\ 
  25 & 2000 & 10000 & 0.8 & -2990 & 28.7 & -2990 & 28.6 & -2993 & 28.2 & -2865 & 28.2 \\ 
   \hline
100 & 200 & 1000 & 0.5 & -660 & 6.7 & -661 & 6.5 & -683 & 8.8 & -292 & 9.2 \\ 
  100 & 200 & 1000 & 0.6 & -660 & 6.7 & -660 & 6.6 & -678 & 8.5 & -286 & 9.0 \\ 
  100 & 200 & 1000 & 0.7 & -660 & 6.7 & -659 & 7.1 & -673 & 7.8 & -282 & 8.0 \\ 
    \hline
100 & 200 & 10000 & 0.5 & -660 & 6.7 & -659 & 6.9 & -682 & 9.1 & -288 & 10.4 \\ 
  100 & 200 & 10000 & 0.6 & -660 & 6.7 & -660 & 5.7 & -678 & 8.8 & -286 & 8.9 \\ 
  100 & 200 & 10000 & 0.7 & -660 & 6.7 & -658 & 6.7 & -674 & 7.3 & -282 & 8.4 \\ 
    \hline
100 & 2000 & 1000 & 0.5 & -3364 & 24.4 & -3364 & 24.8 & -3370 & 27.5 & -2871 & 27.1 \\ 
  100 & 2000 & 1000 & 0.6 & -3364 & 24.4 & -3362 & 24.6 & -3369 & 24.3 & -2868 & 25.3 \\ 
  100 & 2000 & 1000 & 0.7 & -3364 & 24.4 & -3361 & 23.9 & -3371 & 25.6 & -2870 & 25.4 \\ 
  100 & 2000 & 1000 & 0.8 & -3364 & 24.4 & -3362 & 23.9 & -3370 & 26.0 & -2868 & 26.1 \\ 
   \hline
100 & 2000 & 10000 & 0.5 & -3364 & 24.4 & -3362 & 24.0 & -3372 & 25.3 & -2870 & 25.2 \\ 
  100 & 2000 & 10000 & 0.6 & -3364 & 24.4 & -3360 & 24.9 & -3368 & 25.3 & -2867 & 25.4 \\ 
  100 & 2000 & 10000 & 0.7 & -3364 & 24.4 & -3360 & 24.6 & -3370 & 25.5 & -2869 & 25.5 \\ 
  100 & 2000 & 10000 & 0.8 & -3364 & 24.4 & -3362 & 24.5 & -3367 & 24.3 & -2867 & 24.4 \\ 
   \hline
\end{tabular}
\caption{Results of simulation study for effectiveness of estimator
  for log marginal likelihood.}\label{tab:MargLL1}
\end{center}
\end{table}

\begin{table}[hp]\footnotesize
\begin{center}
\begin{tabular}{|llll|rrrr|rrrr|}
  \hline
&&&&\multicolumn{2}{c}{MAPE-GDS}&\multicolumn{2}{c|}{MAPE-LENK}&\multicolumn{2}{c}{Accept
\%}&\multicolumn{2}{c|}{Time (mins)}\\
k & n & M & scale & mean & sd & mean & sd & mean & sd &mean&sd\\
  \hline
 5 & 200 & 1000 & 0.5 & 0.23 & 0.16 & 0.47 & 0.39 & 22.05 & 7.64 & 0.09 & 0.24 \\ 
   5 & 200 & 1000 & 0.6 & 0.11 & 0.12 & 0.52 & 0.45 & 40.46 & 10.82 & 0.02 & 0.04 \\ 
   5 & 200 & 1000 & 0.7 & 0.06 & 0.05 & 0.47 & 0.39 & 57.13 & 8.46 & 0.01 & 0.00 \\ 
    \hline
 5 & 200 & 10000 & 0.5 & 0.17 & 0.08 & 0.50 & 0.37 & 23.99 & 5.32 & 0.04 & 0.01 \\ 
   5 & 200 & 10000 & 0.6 & 0.10 & 0.06 & 0.54 & 0.46 & 40.85 & 7.19 & 0.04 & 0.01 \\ 
   5 & 200 & 10000 & 0.7 & 0.07 & 0.07 & 0.48 & 0.35 & 55.15 & 9.15 & 0.04 & 0.01 \\ 
    \hline
 5 & 2000 & 1000 & 0.5 & 0.02 & 0.01 & 0.07 & 0.05 & 22.14 & 7.65 & 0.04 & 0.07 \\ 
   5 & 2000 & 1000 & 0.6 & 0.01 & 0.01 & 0.06 & 0.06 & 37.76 & 10.34 & 0.03 & 0.08 \\ 
   5 & 2000 & 1000 & 0.7 & 0.01 & 0.01 & 0.06 & 0.05 & 49.55 & 13.39 & 0.01 & 0.00 \\ 
   5 & 2000 & 1000 & 0.8 & 0.01 & 0.01 & 0.04 & 0.02 & 64.60 & 12.93 & 0.01 & 0.00 \\ 
   \hline
 5 & 2000 & 10000 & 0.5 & 0.02 & 0.01 & 0.05 & 0.05 & 25.30 & 6.65 & 0.06 & 0.07 \\ 
   5 & 2000 & 10000 & 0.6 & 0.01 & 0.01 & 0.04 & 0.03 & 36.28 & 7.06 & 0.04 & 0.01 \\ 
   5 & 2000 & 10000 & 0.7 & 0.01 & 0.01 & 0.05 & 0.03 & 51.43 & 14.49 & 0.05 & 0.03 \\ 
   5 & 2000 & 10000 & 0.8 & 0.00 & 0.01 & 0.04 & 0.03 & 71.95 & 11.83 & 0.04 & 0.00 \\ 
   \hline
25 & 200 & 1000 & 0.5 & 0.49 & 0.36 & 0.98 & 0.47 & 2.77 & 2.50 & 0.47 & 0.60 \\ 
  25 & 200 & 1000 & 0.6 & 0.26 & 0.13 & 1.06 & 0.46 & 8.07 & 3.36 & 0.13 & 0.18 \\ 
  25 & 200 & 1000 & 0.7 & 0.18 & 0.10 & 0.80 & 0.43 & 16.16 & 5.08 & 0.06 & 0.11 \\ 
   \hline
25 & 200 & 10000 & 0.5 & 0.52 & 0.25 & 0.93 & 0.65 & 1.67 & 0.87 & 1.15 & 1.99 \\ 
  25 & 200 & 10000 & 0.6 & 0.35 & 0.19 & 1.03 & 0.68 & 6.24 & 3.72 & 0.72 & 1.39 \\ 
  25 & 200 & 10000 & 0.7 & 0.11 & 0.05 & 0.88 & 0.77 & 20.01 & 3.23 & 0.04 & 0.01 \\ 
   \hline
25 & 2000 & 1000 & 0.5 & 0.04 & 0.03 & 0.11 & 0.10 & 2.69 & 1.98 & 0.28 & 0.25 \\ 
  25 & 2000 & 1000 & 0.6 & 0.06 & 0.03 & 0.09 & 0.06 & 4.57 & 3.03 & 0.75 & 1.02 \\ 
  25 & 2000 & 1000 & 0.7 & 0.04 & 0.03 & 0.09 & 0.09 & 15.44 & 9.61 & 0.59 & 1.36 \\ 
  25 & 2000 & 1000 & 0.8 & 0.01 & 0.01 & 0.09 & 0.07 & 43.14 & 12.62 & 0.03 & 0.02 \\ 
   \hline
25 & 2000 & 10000 & 0.5 & 0.10 & 0.05 & 0.08 & 0.06 & 0.75 & 0.68 & 4.34 & 9.53 \\ 
  25 & 2000 & 10000 & 0.6 & 0.07 & 0.03 & 0.09 & 0.05 & 3.65 & 2.76 & 1.97 & 5.95 \\ 
  25 & 2000 & 10000 & 0.7 & 0.03 & 0.03 & 0.08 & 0.06 & 17.07 & 6.33 & 0.42 & 1.45 \\ 
  25 & 2000 & 10000 & 0.8 & 0.01 & 0.01 & 0.10 & 0.08 & 43.27 & 10.27 & 0.05 & 0.01 \\ 
   \hline
100 & 200 & 1000 & 0.5 & 0.27 & 0.23 & 3.50 & 0.82 & 0.32 & 0.29 & 0.49 & 0.85 \\ 
  100 & 200 & 1000 & 0.6 & 0.17 & 0.22 & 2.74 & 0.71 & 0.30 & 0.22 & 1.05 & 3.06 \\ 
  100 & 200 & 1000 & 0.7 & 0.26 & 0.22 & 1.93 & 0.81 & 0.40 & 0.37 & 1.17 & 2.18 \\ 
   \hline
100 & 200 & 10000 & 0.5 & 0.20 & 0.12 & 3.21 & 0.93 & 0.04 & 0.03 & 9.64 & 24.67 \\ 
  100 & 200 & 10000 & 0.6 & 0.22 & 0.14 & 2.62 & 0.75 & 0.08 & 0.07 & 7.62 & 27.89 \\ 
  100 & 200 & 10000 & 0.7 & 0.28 & 0.17 & 2.18 & 0.64 & 0.08 & 0.06 & 1.66 & 1.62 \\ 
   \hline
100 & 2000 & 1000 & 0.5 & 0.06 & 0.05 & 0.24 & 0.14 & 0.34 & 0.38 & 12.03 & 26.31 \\ 
  100 & 2000 & 1000 & 0.6 & 0.04 & 0.04 & 0.19 & 0.12 & 0.60 & 0.53 & 4.61 & 10.84 \\ 
  100 & 2000 & 1000 & 0.7 & 0.07 & 0.04 & 0.23 & 0.13 & 1.10 & 0.88 & 3.66 & 8.46 \\ 
  100 & 2000 & 1000 & 0.8 & 0.06 & 0.03 & 0.21 & 0.12 & 3.16 & 2.26 & 2.74 & 8.74 \\ 
   \hline
100 & 2000 & 10000 & 0.5 & 0.05 & 0.04 & 0.27 & 0.12 & 0.04 & 0.05 & 52.26 & 85.06 \\ 
  100 & 2000 & 10000 & 0.6 & 0.08 & 0.05 & 0.15 & 0.13 & 0.14 & 0.16 & 64.11 & 189.53 \\ 
  100 & 2000 & 10000 & 0.7 & 0.09 & 0.04 & 0.19 & 0.13 & 0.44 & 0.42 & 11.31 & 23.94 \\ 
  100 & 2000 & 10000 & 0.8 & 0.05 & 0.02 & 0.15 & 0.11 & 3.04 & 1.81 & 2.08 & 4.04 \\ 
   \hline
\end{tabular}
\caption{Results of simulation study for effectiveness of estimator
  for log marginal likelihood.}\label{tab:MargLL2}
\end{center}
\end{table}

\section{Practical considerations and limitations of GDS}\label{sec:practical}

This section discusses some practical issues while implementing GDS.  Like the entire body of
MCMC research continues to teach us, more insights
will likely be gleaned as we, and hopefully others, gain additional
experience with the method.  Here, we provide 
some suggestions on how to implement GDS effectively, and mention some
areas in which more research or investigation is needed.

\subsection{Finding the posterior mode and estimating the Hessian}\label{sec:modeHessian}

Searching for the posterior mode is considered, in
general, to be ``good practice'' for Bayesian inference even when
using MCMC; see Step 1 of the ``Recommended Strategy for Posterior
Simulation'' in Section 11.10 of Gelman et. al (2003).  
For small problems, like the example in Section \ref{sec:cauchy}, standard nonlinear optimization algorithms, such as
those found in common statistical packages like R, are sufficient for
finding posterior modes and estimating Hessians. For larger problems,
finding the mode and estimating the Hessian can be more difficult when
using those same tools.  However, there are many different ways to
find the extrema of a function, and some may be more appropriate for
some kinds of problems than for others.  Therefore, one should not
immediately conclude that finding the posterior mode (or modes) is a
barrier to adopting GDS for large or ill-conditioned problems, like
the $29,073$-dimensional model in Section \ref{sec:ads}.

When the log posterior density is smooth and unimodal, a natural
algorithm for finding a posterior mode is one that exploits gradient and
Hessian information in a way that is related to Newton's Method.  Nocedal and Wright (2006) describe many
different nonlinear optimization methods, but most can be classified
as either ``line search'' or ``trust region'' methods.  Line search
methods may be more common; for instance, all of the gradient-based algorithms implemented in the
\texttt{optim} function in R are line search methods. But these methods could be
subject to numerical problems when the log posterior is nearly flat,
or has a ridge, in which case the algorithm may try to evaluate the log posterior at
a point that is so far away from the current value that it generates
numerical overflow.  Trust region methods (Conn, et. al., 2000), on the other hand, tend to be
more stable, because each proposed step is constrained to be within a
particular distance (the ``radius'' of the trust region) of the
current point.  In
short, if one finds that a ``standard'' optimizer for a particular
programming environment is having trouble finding the posterior mode,
there may be other common algorithms that can find the mode more easily.

Neither line-search nor trust-region algorithms necessarily require
explicit expressions for gradients and Hessians, but generating these
structures exactly can also speed up the mode-finding step of GDS.
This approach is in contrast to approximations that use finite
differencing or quasi-Newton Hessian updates.  Of course, one
can always derive the gradient of the log posterior density
analytically, but this can be a tedious process.
We have had success with algorithmic differentiation (AD) software such as
the CppAD library (Bell 2012).  With AD, we need only to write a
function that computes the log posterior density.  The AD library
includes functions that automatically return derivatives of that
function.  The time to compute the gradient of a function is a small
multiple of the time it takes to compute the original function, and
otherwise does not depend on the dimension of the problem.

Estimating the Hessian is useful not only for the mode-finding step,
but also for choosing the covariance matrix of a multivariate proposal
density.  The time it takes for AD software to compute a Hessian can
depend on the dimension of the problem, and working with a dense Hessian
for a large problem can be prohibitively expensive in terms of
computation and memory usage.  However, for many hierarchical models, we assume conditional independence
across heterogeneous units.  For these models, the Hessian of the log
posterior is sparse, with a ``block-diagonal-arrow'' structure
(block-diagonal, but dense on the bottom and right margins).  Thus, we
can achieve substantial computational improvements by exploiting this
sparsity.  The advantage comes in storing the Hessian in a compressed
format, such that zeros are not stored explicitly.  Not only does this
permit estimating larger models on computers with less memory, but it
also lets us use efficient computational routines that exploit that
sparsity. For example, Powell and Toint (1979) and Coleman and More
(1983) explain how to efficiently estimate sparse
Hessians using graph coloring techniques.  Coleman et al. (1985a,
1985b) offer a useful FORTRAN implementation to estimate sparse
Hessians using graph coloring and finite differencing.  Algorithmic
differentiation libraries like CppAD can also exploit
sparsity when computing Hessians.  Both MATLAB and R (through the
Matrix package) can store sparse symmetric matrices in compressed format.

One important consideration is the case of multimodal posteriors.  GDS
does require finding the global posterior mode, and all the models
discussed in this paper have unimodal posterior distributions. When
the posterior is multimodal, one could instead use a mixture of
normals as the proposal distribution.
The idea is to not only find the global mode, but any local ones as
well, and center each mixture component at each of those local
modes. The GDS algorithm itself remains unchanged, as long as the global
posterior mode matches the global proposal mode. 

We recognize that finding all of the local modes could be a hard
problem, and there is no
guarantee that any optimization algorithm will find all local
extrema.  But, by the same token, this problem can be resolved efficiently in a multitude of complex Bayesian statistical 
models if one
uses the correct tools.  And it is only a matter of time before these
tools are more widely available in standard statistical programming
languages like R.
The nonlinear optimization literature is rife with methods that help facilitate
efficient location of multiple modes, even if there is no guarantee of
finding them all.  Also, note that
even though MCMC sampling chains are, in \emph{theory}, guaranteed
to explore the entire space of any posterior distribution (including
multiple regions of high posterior mass), there is no guarantee that
this will happen after a large finite number of iterations for
general nonconjugate hierarchical models.  Other estimation
algorithms that purport to be robust to multimodal posteriors offer no
such guarantees either.

\subsection{Choosing a proposal distribution}

Like many other methods that collect random samples from posterior
distributions, the efficiency of GDS depends in part on a prudent selection
of the proposal density $g(\theta)$.  For the examples in this paper,
we used a multivariate normal density that is centered at the posterior
mode, with a covariance matrix that is proportional to the inverse of
the Hessian at the mode.  One might then wonder if there is an optimal
way to determine just how ``scaled out'' the proposal covariance needs to
be.  At this time, we think that trial and error is, quite frankly,
the best alternative.  For example, if we start with a small $M$ (say,
100 draws), and find that $\Phitheta>1$ for any of the $M$ proposals,
we have learned that the proposal density is not valid, at little
computational or real-time cost.  We can then re-scale the proposal
until $\Phitheta<1$, and then gradually increase $M$ until we get a
good approximation to $p(u)$.  In our experience, even if an
acceptance rate appears to be low (say, 0.0001), we can still collect
draws in parallel, so the ``clock time'' remains much less than the
time we spend trying to optimize selection of the proposal.

For example, in the Cauchy example in Section \ref{sec:cauchy}, we set
the proposal covariance to be the inverse Hessian at the posterior
mode, scaled by a factor of 200.  We needed such a large scale factor because the normal
approximation at the mode shows no correlation, even though there is
obvious correlation in the tails.  If one knew
upfront the extent of the tail dependence, one might have chosen a
proposal density that is more highly correlated, and that might 
give a higher acceptance rate.  But of course one seldom, if ever, knows the
shape of any target posterior density up front.  So even though an acceptance percentage of 1.3\% may appear
to be low, we should consider the amount of time it would take to
improve the proposal density, and especially the number of MCMC
iterations it would take to get enough draws that are equivalent to
the same number of independent GDS draws.

\subsection{Cases requiring further research}
This paper demonstrated that GDS is a viable alternative to MCMC for a large
class of Bayesian non-Gaussian and Gaussian hierarchical models. Of course it would be myopic to claim that GDS
is appropriate for all models. By the same token, we cannot assert that GDS would not work
for any of the models described below.  These models are topics requiring additional research.

\paragraph{Models with discrete or combinatorial optimization elements}  In
models that include both discrete and continuous parameters, finding the
posterior mode becomes a mixed-integer nonlinear program (MINLP).  An
example is the Bayesian variable selection problem (George and McCulloch
1997).  The difficulty lies in the fact that MINLPs are known to be NP-complete, and thus may not scale
well for large problems.  Hidden Markov
models with multiple discrete states might be similarly difficult to
estimate using GDS.  Also, it is not immediately clear how one might
select a proposal density when some parameters are discrete.

\paragraph{Intractable likelihoods or posteriors}  There are many
popular models, namely binary, ordered and multinomial probit models, for which the
likelihood of the observed data is not available in closed form.  When
direct numerical approximations to these likelihoods (e.g., Monte
Carlo integration) is not tractable, MCMC with data
augmentation is a popular estimation tool (e.g., Albert and Chib 1993).
That said, recent advances 
in parallelization using graphical processing units (GPUs) might make numerical estimation of integrals
more practical than it was even
10 years ago; see Suchard et al. (2010).  If this is the case, and the
log posterior remains sufficiently
smooth, then GDS could be a viable, efficient alternative to data
augmentation in these kinds of models.  

\paragraph{Missing data problems}  MCMC-based approaches to multiple
imputation of missing data could suffer from the same kinds of
problems:  the latent parameter, introduced
for the data augmentation step, is only weakly identified on its own.
Normally, we are not interested in the missing values themselves. If
the number of missing data points is small, perhaps one could treat the
representation of the missing data points as if they were parameters.
But the implications of this require additional research.

\paragraph{Spatial models, and other models with dense Hessians}
GDS does not
explicitly require conditional independence, so one might consider
using it for spatial or contagion models (e.g., Yang and
Allenby 2003).  However, without a conditional independence
assumption, the Hessian of the log posterior will not be sparse, and
that may restrict the size of datasets for which GDS is practical.

\section{Conclusions}
In this paper, we presented a new method, Generalized
Direct Sampling (GDS), to sample from posterior distributions. This method has the potential to bypass MCMC-based Bayesian inference for
large, complex models with continuous, bounded posterior densities.  Unlike MCMC, GDS generates independent draws that one could collect in
parallel. The implementation of GDS is straightforward, and requires
only a function that returns the value of the unnormalized log
posterior density.  In addition, GDS allows for fast and accurate
computation of marginal likelihoods, which can then be used for model
comparison.  

There are many other ways to conduct Bayesian inference,
and continued improvement of MCMC remains an important stream of
research.   Nevertheless, it would be hard to ignore the opportunities for
parallelization that make algorithms like GDS very attractive alternatives.  Of
course, one could employ parallel computational techniques
as part of a sequential algorithm.  But, to repeat an earlier sentence, using parallel technology to generate a single draw is not the same as generating all of the required draws themselves in parallel. By exploiting the advantages of parallel computing, as in this paper, GDS could prove to be a successful addition to the Bayesian practitioner's computational toolkit.

\begin{singlespace}

\end{singlespace}

\end{document}